# Electrical properties of inhomogeneous tungsten carbide Schottky barrier on 4H-SiC

Vivona M[1], Greco G[1], Bellocchi G[2], Zumbo L[2], Di Franco S[1], Saggio M[2], Rascunà S[2] and Roccaforte F[1]

[1]CNR-IMM, Strada VIII n.5 Zona Industriale, I-95121 Catania, Italy
[2] STMicroelectronics, Stradale Primosole 50, I-95121 Catania, Italy

E-mail: marilena.vivona@imm.cnr.it

**Abstract**

In this paper, the electrical behavior of tungsten carbide (WC) Schottky barrier on 4H-SiC was investigated. First, a statistical current-voltage (I-V) analysis in forward bias, performed on a set of equivalent diodes, showed a symmetric Gaussian-like distribution of the barrier heights after annealing at 700°C, where a low Schottky barrier height ($\Phi_B$=1.05 eV) and an ideality factor n=1.06 were measured. The low value of the barrier height makes such a WC contact an interesting candidate to reduce the conduction losses in 4H-SiC Schottky diodes. A deeper characterization has been carried out, by monitoring the temperature dependence of the I-V characteristics and the behavior of the relevant parameters $\Phi_B$ and n. The increase of the barrier height and decrease of the ideality factor with increasing temperature indicated a lateral inhomogeneity of the WC/4H-SiC Schottky contact, which was described by invoking the Tung's model. Interestingly, the temperature dependence of the leakage current under reverse bias could be described by considering in the thermionic field emission model the temperature dependent barrier height related to the inhomogeneity. These results can be useful to predict the behavior of WC/4H-SiC Schottky diodes under operative conditions.

## 1. Introduction

Today, hexagonal silicon carbide (4H-SiC) is recognized as material of choice to overcome the limits of conventional silicon technology, in terms of power losses reduction and energy efficiency improvement [1]. This relies on the superior physical properties of the 4H-SiC compared to Si, like a wider band gap (3.2 eV), higher breakdown electrical field (3MV/cm), saturation electron velocity ($2\times10^7$ cm/s) and thermal conductivity (3.7 W·K$^{-1}$·cm$^{-1}$) [2].

Among the 4H-SiC devices, Schottky barrier diodes (SBDs) have reached a mature technological level with a variety of products commercially available by several companies, providing low on-state voltage drop, high breakdown voltage, high switching speed, combined with possibility to operate at high temperature [3].

Since the core of a SBD is the metal/semiconductor interface, in the last two decades several studies have investigated the electrical properties of various metallization schemes for Schottky contacts to 4H-SiC and the current transport through these interfaces [4,5,6]. Nowadays, in state-of-the-art 4H-SiC commercial devices, the Schottky contact is typically based on titanium (Ti) or molybdenum (Mo), as these metals enable to obtain low Schottky barrier height values (in the range 1.1-1.2eV) and can be easily integrated in the manufacturing steps. However, since a further reduction of the conduction losses in SBDs is highly desired in many applications, the reduction of the barrier height (i.e., the device turn-on voltage) is one of the current challenges in 4H-SiC SBD technology [7].

In literature, metallization schemes based on low-work-function metals with high melting point (W, Mo, ...) [8,9,10,11,12,13] or tunable composition have been investigated to optimize the performance of 4H-SiC Schottky diodes. As an example, Stöber *et al.* [12] proposed the use of molybdenum nitride ($MoN_x$) Schottky contacts, controlling the barrier height by varying the nitrogen fraction in the reactive sputtering metal deposition. On the other hand, by thermal annealing of a thin tungsten (W) layer, Knoll *et al.* [13] observed the formation of tungsten carbide ($W_2C$), with a smooth interface morphology, high thermal stability and low turn-on voltage in 4H-SiC diodes. In this context, only few papers studied the tungsten carbide barrier material, being most of them limited to structural, morphological and optical characterization of this contact on 6H-SiC [14,15,16].

The electrical characterization of a contact relies on the study of the Schottky barrier height $\Phi_B$, which governs the forward voltage drop, and the reverse leakage current. Peculiarity of an ideal contact is featuring a barrier height independent of temperature, voltage and measurement method, with the $\Phi_B$ value close to that predicted by the Schottky-Mott rule [17]. In contrast, in "real" cases, Schottky contacts on 4H-SiC can exhibit deviations from the ideal behavior, due to interface inhomogeneities related to surface/interface states, defects, processing contaminations, etc. [3].

Several studies have discussed the inhomogeneity of Schottky contacts on 4H-SiC based on different metals (Ti, Ni, $Ni_2Si$, W, ....) [18,19,20,21,22,23,24]. These studies often invoked the Tung's model [25], which assumes a local lateral inhomogeneity of the Schottky barrier height. Specifically, nanometer-size regions characterized by a low barrier height value and embedded in a uniform high barrier background, interact one another with a resulting pinch-off effect on the low barrier height regions [25]. However, the electrical behavior and the homogeneity of tungsten carbide (WC) Schottky contacts on 4H-SiC are still unexplored.

This study reports on the electrical behavior of WC/4H-SiC Schottky contacts. In particular, a temperature dependent current-voltage characterization (I-V-T) enabled to get insights on the inhomogeneous nature of the barrier and to explain the forward and reverse conduction mechanisms through the inhomogeneous WC/4H-SiC interface.

## 2. Experimental details

Commercial n-type 4H-SiC wafer, with a 9.5 μm thick epitaxial layer having a nominal doping concentration $N_D=8\times10^{15}cm^{-3}$, was used as starting material in our study. On this sample, Schottky diodes with an active contact area $A=4.53\times10^{-2}$ cm$^2$ have been fabricated, using tungsten carbide (WC) as barrier metal. Before the front-side processing of the wafer, a large-area back-side contact has been fabricated by Ni deposition followed by rapid thermal annealing (RTA) at 950 °C in $N_2$ [26]. Then, 80nm-thick WC layer was deposited by DC magnetron sputtering, and defined by optical lithography and lift-off process. The relevant electrical parameters of the contacts (ideality factor n and Schottky barrier height $\Phi_B$) were determined by means of current-voltage (I-V) measurements on a set of 40 equivalent diodes, performed in a Karl-Suss MicroTec probe station equipped with a parameter analyzer. The Schottky diodes were characterized both before (as-deposited) and after annealing treatment at 700 °C for 10 min in $N_2$-atmosphere. For selected samples, the temperature-dependence of both forward and reverse I-V characteristics (I-V-T) was monitored in the range 25-125°C, to get more insights on the barrier inhomogeneity and leakage mechanisms.

## 3. Results and Discussion

Firstly, in order to evaluate the uniformity of the WC/4H-SiC Schottky barrier formation, the distribution of the ideality factor n and barrier height $\Phi_B$ was derived from forward I-V characteristics acquired on a set of equivalent diodes fabricated on the wafer, in the as-deposited and annealed samples. Fig.1 reports in a semilog plot the I-V characteristics representative of the average behavior of the as-deposited and annealed WC/4H-SiC tested diodes.

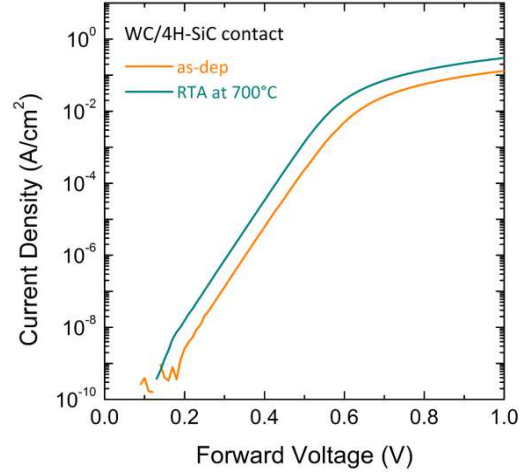

Fig.1. Representative forward I-V characteristics of WC/4H-SiC Schottky diodes for the as-deposited contact and after thermal annealing at 700°C.

Evidently, the forward I-V curves present a wide linearity region over several decades. Above a certain bias (0.6-0.8V), the I-V curves start to bend, due to the dominant contribution of the series resistance. After annealing at 700°C, the forward I-V curves of the WC/4H-SiC diodes exhibit a slight negative shift with respect to the as-deposited case, suggesting a reduction of the barrier height.

By applying the thermionic emission (TE) model [17], the ideality factor n and the Schottky barrier height $\Phi_B$ were derived by fitting the linear region of the semilog forward I-V curves, according to the relation:

$$I = AA^*T^2 \exp\left[\left(-\frac{q\phi_B}{k_BT}\right)\right] \exp\left[\left(\frac{qV_F}{nk_BT}\right)\right] \quad (1)$$

where A is the diode area, $A^*$ is the theoretical Richardson constant of 4H-SiC (146 A·cm$^{-2}$·K$^2$) [20], $k_B$ is the Boltzmann constant, q is the electron charge, $V_F$ is the voltage applied across the metal/semiconductor interface and T is the absolute temperature.

Fig.2 shows the statistical distribution of the barrier height $\Phi_B$ values for the as-deposited and annealed contacts, determined on a set of 40 diodes in different positions of the wafer. Notably, narrow distributions of the $\Phi_B$ values were obtained, thus indicating a good reproducibility of the WC barrier formation process. In the as-deposited contact, an asymmetric distribution of the measured barrier heights was observed, with the barrier value peaked at 1.11-1.12 eV. On the other hand, with the annealing treatment at 700°C, the barrier heights distribution became more symmetric. In particular, the barrier heights distribution can be assimilated to a Gaussian curve (Fig.2b) centered at 1.075 eV and with a standard deviation of 0.013eV.

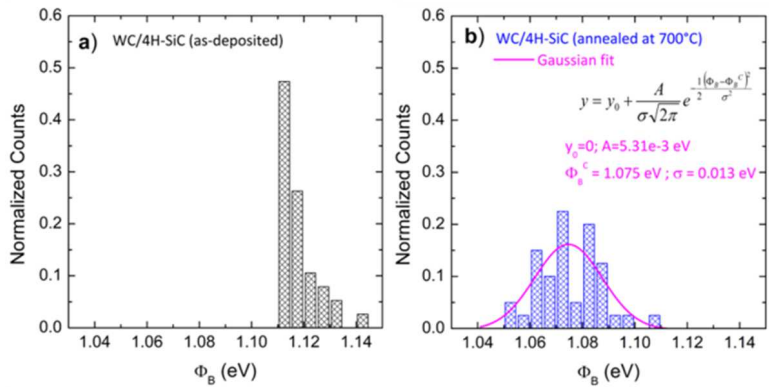

Fig.2. Statistical distribution of the measured barrier heights in WC/4H-SiC Schottky contacts, extracted from I-V measurements on a set of diodes in the as-deposited sample (a) and after thermal annealing at 700°C (b).

Fig.3 reports the average values of the ideality factor and Schottky barrier height. The error bars are associated to the statistical distribution of the results obtained in the measured diodes over the sample. Regarding the ideality factor, independent of the annealing treatment, a nearly ideal behavior with n=1.03 is observed. On the other hand, the average barrier height $\Phi_B$ decreases from 1.12 eV in the as-deposited contact to 1.075 eV after annealing at 700 °C. Noteworthy, the errors associated to the measurements of ideality factor and barrier height are in the order of 0.7-2 %, thus indicating a good stability of the contact properties upon annealing treatment.

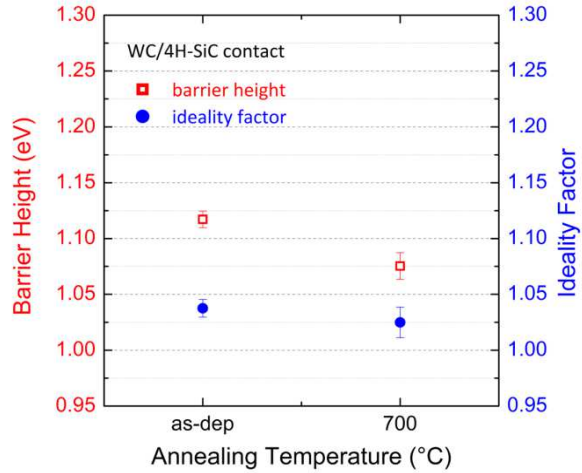

Fig.3. Average values of ideality factor n and barrier height $\Phi_B$ determined from the forward I-V characteristics of WC/4H-SiC Schottky diodes for the as-deposited and 700°C annealed contacts.

According to the Schottky–Mott rule [27,28], the ideal barrier height $\Phi_B^{ideal}$ can be predicted based on the relative energy band alignment. Specifically, for an ideal Schottky contact to n-type semiconductor $\Phi_B^{ideal}$ is given by the difference between the metal work-function ($\Phi_M$) and the semiconductor electron affinity ($X_{SiC}$), i.e., $\Phi_B^{ideal}=\Phi_M-X_{SiC}$. Considering $X_{SiC}$=3.2eV for 4H-SiC [29], an ideal barrier height $\Phi_B^{ideal}$=1.7eV is expected for the WC/4H-SiC contact, taking into account a work function of 4.9 eV for WC [30]. Clearly, the experimental data determined by the I-V characterization do not match with the theoretical calculations. In fact, in a "real" metal/semiconductor contact the presence of surface states (related to roughness, surface contaminants, residual thin interfacial oxide layers, etc.) is responsible for a deviation from the Schottky–Mott rule [17]. Moreover, the metal work function can be also affected by the deposition technique and/or by the annealing treatments.

For a deeper investigation of the Schottky contact behavior, a temperature-dependent current-voltage characterization (I-V-T) was carried on this system. Such a characterization gives information on the local degree of homogeneity of the barrier and enables to elucidate the current transport mechanisms through the metal/4H-SiC interface. These measurements were performed on one of the contacts annealed at 700 °C. The forward J-V characteristics (with J the current density) are displayed in Fig.4 for five different measurement temperatures (from 25 to 125 °C). As predicted by thermionic emission model, the diode forward current increases with increasing the measurement temperature.

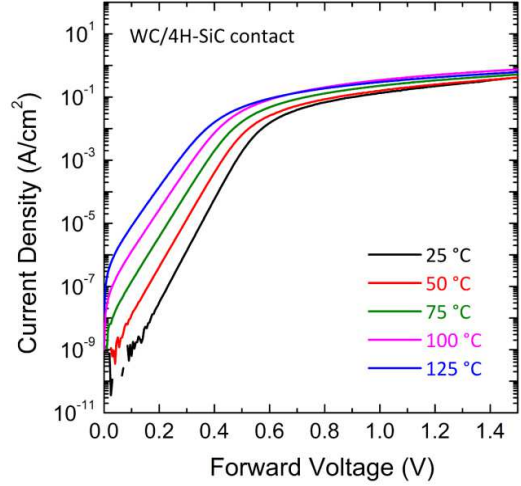

Fig.4. Forward J-V characteristics acquired at different temperatures ranging from 25 to 125°C with step=25°C for the WC/4H-SiC Schottky contact annealed at 700°C.

The ideality factor n and barrier height $\Phi_B$ determined from these curves are reported in Fig.5 as function of the measurement temperature. As can be seen, while the barrier height increases with increasing temperatures, the ideality factor decreases. The temperature dependence of ideality factor n and barrier height $\Phi_B$ indicates that the WC/4H-SiC Schottky contact is characterized by a certain degree of local inhomogeneity. Such a behavior was described by applying the Tung's theory [20].

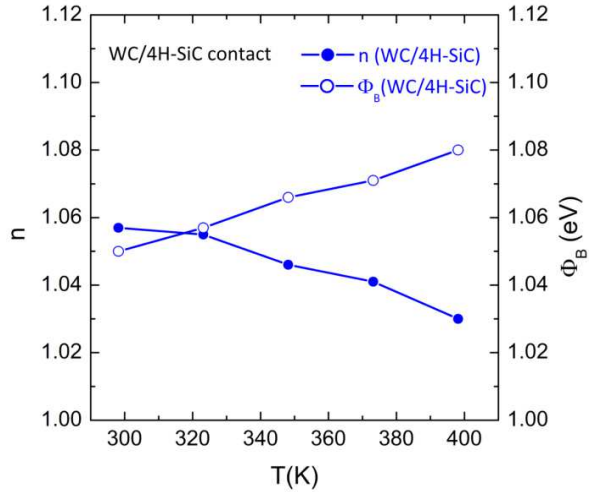

Fig.5. Temperature-dependence of ideality factor n and barrier height $\Phi_B$ for the WC/4H-SiC Schottky contact subjected to thermal annealing at 700°C.

Accordingly, the inhomogeneity of the metal/semiconductor interface can be explained by considering a distribution of regions (patches) characterized by different low barrier heights and different area, embedded in an ideal high-barrier background. Hence, at lower temperature, the current transport will be strongly influenced by the contact inhomogeneity and a lower barrier is measured. At higher temperature, the transport will approach the ideal thermionic behavior and a higher barrier is measured. However, the expression of the forward current given in Eq. (1) does not take into account the experimentally observed temperature dependence of n and $\Phi_B$. Hence, the temperature dependence of the saturation current $I_S$ derivable from Eq. (1) and expressed as

$$I_S = AA^*T^2 \exp[(-q\phi_B/k_BT)] \qquad (2)$$

allows to extrapolate the "effective" values of the barrier height and of the Richardson's constant as slope and y-intercept of a conventional "Richardson's plot" $\ln \frac{I_S}{T^2}$ vs $\frac{q}{k_B T}$.

Such a plot, reported in Fig.6, provides an effective barrier height $\Phi_B^{eff}$=0.96 eV, which can be regarded as an average barrier of the patches and is lower than the barrier height derived by the forward I-V curve at room temperature (1.05 eV). On the other hand, an effective Richardson's constant $A^*_{eff}$=4.7 A·cm$^{-2}$·K$^{-2}$ can be estimated, which is two orders of magnitude lower that the theoretical value (146 A·cm$^{-2}$·K$^{-2}$). Since the y-intercept enables to determine the product AA*, this results indicates that the effective active area interested by the current transport is smaller than the entire area of the contact, as assumed in the application of the TE model.

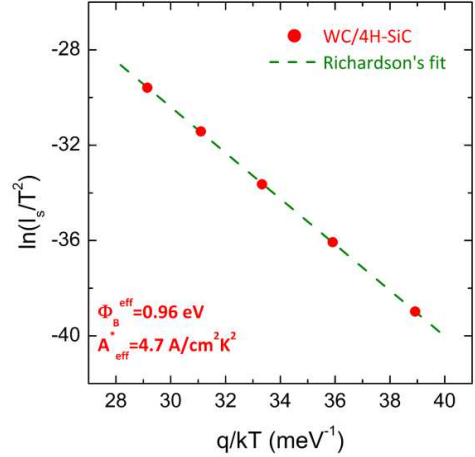

Fig.6. Richardson's plot ln(Is/T$^2$) *vs* 1/k$_B$T for determining the effective barrier height $\Phi_B^{eff}$ and Richardson's constant $A^*_{eff}$ in the WC/4H-SiC Schottky contact annealed at 700°C.

In order to quantify the Schottky barrier inhomogeneity, it is useful to compare the plot nkT *vs* kT with the ideal case (n=1), as depicted in Fig.7a. As can be seen, the experimental data can be fitted by a straight line parallel to that of the ideal case, allowing to express the ideality factor as $n = 1 + T/T_0$ and calculate the so-called "T$_0$ anomaly", resulting T$_0$=34K for our case. This expression for the ideality factor agrees with the Tung's formalism for the case where the inhomogeneity can be described with a Gaussian distribution of effective barrier heights related to the low-barrier patches [31]..

According to Tung's approach, the effective barrier $\Phi_{Bi}^{eff}$ of a single low-barrier patch is related to the surrounding uniform high barrier $\Phi_B^0$ background by:

$$\Phi_{B_i}^{eff} = \Phi_B^0 - \gamma_i \left(\frac{V_{bb}}{\eta}\right)^{1/3} \quad (3)$$

where V$_{bb}$ is the band bending and $\eta = \varepsilon_S/qN_D$. The parameter γ takes into account the patch characteristics through the relation $\gamma = (3 \Delta R_0^2/4)^{1/3}$, where $\Delta$ is the local deviation from the ideal barrier $\Phi_B^0$ and R$_0$ is the patch radius (assuming a circular geometry) [31]. In general, an inhomogeneous Schottky contact is characterized by a Gaussian distribution of γ$_i$ parameters, whose standard deviation σ$_\gamma$ is related to the standard deviation of the effective barrier distribution $\sigma_{\phi_{eff}}$ by the relations [31]:

$$T_0 = q \sigma_\gamma^2 /3k_B \ \eta^{2/3} V_{bb}^{1/3} = q \sigma_{\phi_{eff}}^2/3 k_B V_{bb} \quad (4)$$

Hence, considering the value T$_0$= 34 K obtained in our case (Fig. 7a), a standard deviation $\sigma_{\phi_{eff}}$=0.12 eV can be derived. Furthermore, in an inhomogeneous Schottky contact, the ideal barrier height $\Phi_B^0$ can be determined as the limit to the ideal behavior (n=1) of the so-called Schmitsdorf's plot (barrier height $\Phi_B$ *vs* ideality factor n) [31]. In particular, from such a plot, reported in Fig.7b, it was possible to determine an ideal barrier height $\Phi_B^0$=1.11 eV for the WC/4H-SiC contact. This value corresponds to the uniform high barrier background surrounding the patches.

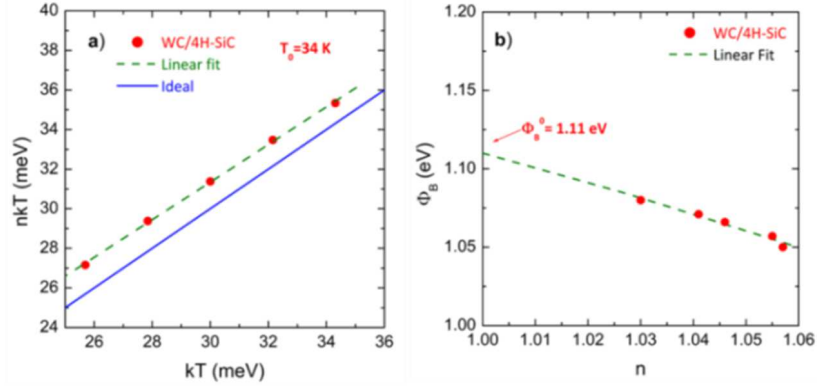

Fig.7. a) nkT vs kT plot expressing the deviation from the ideal case (depicted as solid line), showing the $T_0$ anomaly; b) $\Phi_B$ vs n plot from which the ideal homogeneous $\Phi_B^0$ (at n=1) is extrapolated for the WC/4H-SiC Schottky contact annealed at 700°C.

Therefore, it can be concluded that our inhomogeneous WC/4H-SiC Schottky contact can be described by a Gaussian distribution of low barrier patches with an average effective barrier $\Phi_B^{eff}$=0.96eV and standard deviation of $\sigma_{\phi_{eff}}$=0.12 eV, embedded in a uniform ideal barrier $\Phi_B^0$=1.11eV

In order to get additional insights on the barrier height and current conduction mechanism, the WC/4H-SiC contact was studied also under reverse bias, monitoring the temperature-dependence of the reverse I-V characteristics of the diodes. It is known that the presence of interface inhomogeneities, whose presence was previously observed under forward bias for the WC/4H-SiC interface, can enhance the tunneling contribution to the reverse leakage current [32,33]. This behavior can be described by the thermionic field emission (TFE) model [34]. In particular, under reverse bias, SiC-diodes show a very high electric field in the depletion region (about 10 times higher than the conventional Si-diodes). Such a high electric field entails a thinning of the potential barrier, thus enabling the electrons to tunnel through with an increase of the leakage current by many orders of magnitude [35]. Therefore, while in Si diodes the reverse current is typically described by the thermionic emission model accounting the image force lowering, the reverse characteristics of 4H-SiC Schottky diodes is better described considering an additional tunneling contribution, that is, using the TFE mechanism [36,37]. However, there is no consensus in literature on the model describing the reverse characteristics of 4H-SiC Schottky diodes. In fact, while some papers uses the classical TFE model [34,38], whose validity depends on doping concentration and bias range [39], other works include the image force lowering effect in the TFE model to fit the experimental data [37,40].

In particular, taking into account the doping level and the reverse bias range of our case, the TFE (without force lowering effect) should be the most appropriate model to describe the reverse characteristics.

According to the TFE model, the current density J can be expressed as:

$$J = A^* T^2 \sqrt{\frac{q \pi E_{00}}{kT}} \sqrt{V_R + \frac{\phi_B}{\cosh\left(\frac{qE_{00}}{kT}\right)^2}} \exp\left(-\frac{\phi_B}{E_1}\right) \exp\left(\frac{V_R}{E_2}\right) \quad (5)$$

where $E_{00} = (h/4\pi)\sqrt{N_D/m^* \varepsilon_{SiC}}$, $E_1 = E_{00} \times \tanh(qE_{00}/kT)^{-1}$ and $E_2 = E_{00} \times \left(qE_{00}/kT - \tanh(qE_{00}/kT)\right)^{-1}$ and with h the Planck constant, m* the effective mass of electron and $\varepsilon_{SiC}$ the dielectric constant of the semiconductor.

Firstly, a barrier height value of 1.11 eV was considered in Eq. (7), which corresponds to the ideal barrier obtained by Schmitsdorf's plot of Fig. 7b. However, under this assumption a significant discrepancy between the experimental reverse characteristics and the simulated curves was found for all temperatures. Similar discrepancy was found by considering the effective barrier height of 0.96 eV obtained by the Richardson's plot of Fig. 6.

Hence, the barrier inhomogeneity was included in the TFE model, by considering in Eq. (7) the experimental temperature-dependence of the barrier height. Fig.8 displays the experimental J-V reverse characteristics, measured in the WC/4H-SiC contact annealed at 700°C, together with the simulated curves according to TFE model including the barrier inhomogeneity. As can be seen in Fig. 8, a good agreement with experimental curves is observed when the barrier inhomogeneity is included in the TFE model. Specifically, the temperature-dependence of the barrier, with values derived from the analysis of forward I-V characteristics of Fig.4, is considered in the simulated curves reported by dashed lines in Fig. 8.

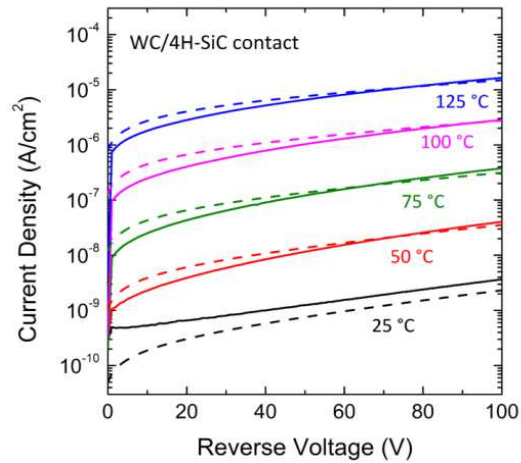

Fig.8. Experimental reverse J-V characteristics (solid-line) for WC/4H-SiC contact annealed at 700°C compared with curves simulated by applying the TFE model including the barrier inhomogeneity (dashed-line).

From our approach, we can conclude that the temperature dependence of the barrier height, as observed in the forward I-V-T curves, must be taken into account to better describe the current transport at the WC/4H-SiC by the TFE model under reverse bias.

## 4. Conclusions

In this work, the electrical properties of the tungsten carbide (WC) Schottky barrier on 4H-SiC have been studied. In particular, based on a set of measurements on several diodes, the contacts showed a nearly ideal behavior (n=1.06) and a narrow distribution of the barrier height values upon annealing at 700°C.

The temperature dependent I-V characterization of the annealed contact revealed an increase of the barrier height and a decrease of the ideality factor with increasing measurement temperature, thus indicating a lateral inhomogeneity of the WC/4H-SiC Schottky contact. This behavior was described with the Tung's model on inhomogeneous barriers. Noteworthy, the temperature dependence of the reverse leakage current could be well described by including the temperature dependent barrier height, measured in forward bias, in the classic thermionic field emission formalism.

The low value of the barrier height (1.05eV) makes such a WC contact an interesting candidate to reduce the conduction losses in 4H-SiC Schottky diodes. Moreover, the results are useful to predict the behavior of WC/4H-SiC Schottky diodes under operative conditions.

## Acknowledgements

The authors would like to acknowledge F. Giannazzo (CNR-IMM) for fruitful discussions on these experiments and related results. Part of this research activity has been carried out in the framework of the European ECSEL JU project REACTION (grant agreement n. 783158).